\documentclass[iop]{emulateapj}
\usepackage{apjfonts}
\usepackage{graphicx,bm,tikz,amsmath,ragged2e}
\begin{document}

\title{On the R-Process Enrichment of Dwarf Spheroidal Galaxies}
\author{Joseph Bramante$^1$, Tim Linden$^{2}$}
\affil{$^1$ Department of Physics, 225 Nieuwland Science Hall,\\ University of Notre Dame, Notre Dame, IN 46556, USA}
\affil{$^2$ Center for Cosmology and AstroParticle Physics (CCAPP) and Department of Physics \\ The Ohio State University, Columbus OH, 43210, USA}
\shortauthors{Bramante \& Linden}

\begin{abstract}
Recent observations of Reticulum II have uncovered an overabundance of r-process elements, compared to similar ultra-faint dwarf spheroidal galaxies (UFDs). Because the metallicity and star formation history of Reticulum II appear consistent with all known UFDs, the high r-process abundance of Reticulum II suggests enrichment through a single, rare event, such as a double neutron star (NS) merger. However, we note that this scenario is extremely unlikely, as binary stellar evolution models require significant supernova natal kicks to produce NS-NS or NS-black hole mergers, and these kicks would efficiently remove compact binary systems from the weak gravitational potentials of UFDs. We examine alternative mechanisms for the production of r-process elements in UFDs, including a novel mechanism wherein NSs in regions of high dark matter density implode after accumulating a black-hole-forming mass of dark matter. We find that r-process proto-material ejection by tidal forces, when a single neutron star implodes into a black hole, can occur at a rate matching the r-process abundance of both Reticulum II and the Milky Way. Remarkably, dark matter models which collapse a single neutron star in observed UFDs also solve the missing pulsar problem in the Milky Way Galactic center. We propose tests specific to dark matter r-process production which may uncover, or rule out, this model.
\end{abstract}

\maketitle


The Dark Energy Survey has discovered a new dwarf spheroidal galaxy, named Reticulum II, which lies at a distance of only $\sim$30~kpc from Earth~\citep{Bechtol:2015cbp, Koposov:2015cua}. The proximity of Reticulum II benefits indirect searches for dark matter (DM) annihilation. In fact, an analysis of data from the Fermi Large Area Telescope found a tentative (2.4 -- 3.2$\sigma$ local significance) excess in GeV $\gamma$-rays emanating from the position of the Reticulum II dwarf~\citep{Drlica-Wagner:2015xua, Geringer-Sameth:2015lua, Hooper:2015ula}. Spectroscopic measurements of Reticulum II stars identify Reticulum II as an ultrafaint dwarf spheroidal galaxy (UFD), with total stellar luminosity of only 2360$\pm$200 L$_\odot$ and a mass to light ratio of 470$\pm$210~\citep{Simon:2015fdw}, making it an excellent target for DM indirect detection studies~\citep{Simon:2015fdw, Bonnivard:2015tta}. 

Recent measurements of stellar spectra indicate that Reticulum II is unique among known UFDs. The majority of stars in Reticulum II are overabundant in elements heavier than Zinc, a signature of rapid neutron capture (aka r-process) enrichment~\citep{2015arXiv151201558J, 2016arXiv160104070R}. This is intriguing, because nine similar UFDs, namely Segue 1, Hercules, Leo IV, Segue II, Canes Venatici II, Bootes I, Bootes II, Ursa Major II, and Coma Berenices, show only trace r-process enrichment~\citep{2015arXiv151007632J}, which may be consistent with r-process materials accreted through interactions of these systems with the Galactic disk. Moreover, the s-process metallicity of Reticulum II, [Fe/H]~=~-2.65$^{+0.07}_{-0.07}$~\citep{Simon:2015fdw} is consistent with, and even lower than, other UFDs~\citep{Kirby:2008ab, 2014ApJ...794...89K, 2014ApJ...786...74F, 2015arXiv151007632J}. Two of the nine stars observed by~\citet{2015arXiv151201558J} show no r-process enrichment, a possible indication of multiple star formation epochs.

The excess of neutron-rich elements in Reticulum II has implications for the production of heavy r-process elements, which occurs at astrophysical sites harboring copious free neutrons. Within recent decades, core collapse supernovae have been considered as a source of r-process materials \citep{1986ApJ...309..141D,MeyerMathews1992ApJ...399..656M}. However, r-process elements in Milky Way (MW) stars show characteristic peak abundances at atomic masses $A = 80, 130, 195$ \citep{1957RvMP...29..547B}, and the entropy provided by core-collapse supernovae appears too low to reproduce the third peak \citep{Qian:1996xt,Thompson:2001ys,Fischer:2009af}. Alternatively, decompressing neutron fluid ejected from, e.g., the merger of a neutron star (NS) with another NS or a black hole (BH), reproduces the $A = 195$ peak ~\citep{1976ApJ...210..549L,1977ApJ...213..225L,1989Natur.340..126E,1994ApJ...431..742D,1999ApJ...525L.121F}, and could be the source of most r-process elements \citep{1982ApL....22..143S,Arnould:2007gh,Surman:2008qf,2015ApJ...807..115S, 2014MNRAS.438.2177M, 2015MNRAS.447..140V, 2015A&A...577A.139C}. More recent proposals for r-process production include the accretion induced collapse of white dwarfs~\citep{1992ApJ...391..228W} and rapidly rotating magnetars~\citep{2008ApJ...676.1130M}. 

Another method to distinguish between r-process models is to utilize the stochasticity of r-process enrichment in small, isolated, metal poor systems (e.g.~UFDs). For example, lest they overproduce r-process elements, frequent iron core collapse supernovae must each produce a small r-process abundance ($\lesssim 10^{-7}~{M_{\rm \odot}}$), and also in this case r-process abundance should scale with stellar metallicity. On the other hand, double NS mergers are expected to be rare and so they can produce copious r-process materials ($\gtrsim 10^{-4}~{M_{\rm \odot}}$). Thus, the observation of significant r-process enrichment in Reticulum II (alongside its typical metallicity for an ultrafaint dwarf spheroidal galaxy), and the lack of r-process enrichment in any other UFD, points towards a rare event, like a double NS merger, for the r-process enrichment of Reticulum II~\citep{2015arXiv151201558J}. However, as we will show, NS mergers in UFDs appear too rare to account for the enrichment of Reticulum II. 

In this article we examine the viability of many r-process production models in light of the r-process abundances now observed in both UFDs and the MW. We also propose a new r-process production site: the neutron-rich fluid ejected from DM-induced implosions of NSs. We find that DM-induced NS implosions could account for both the r-process abundance observed in Reticulum II and the MW. In Section \ref{sec:nsms} we study the rate of NS mergers in UFDs and find this scenario for r-process production to be disfavored at $>$3$\sigma$. Section \ref{sec:alts} examines alternative proposals for r-process production. DM-imploded NSs are introduced as an r-process production site in Section \ref{sec:darkrprocess}, and DM r-process enrichment of UFDs and the MW is compared to other proposals in Section \ref{sec:darkrates}. Section \ref{sec:rglob} notes that a comparison between the r-process enrichment of UFDs and that of globular clusters can provide a diagnostic test specific to DM dominated r-process production. In Section \ref{sec:conc}, we conclude with additional applications of r-process observations to studies of DM, primordial black holes, and by extension the primordial power spectrum.

\section{Neutron Star Mergers in UFDs}
\label{sec:nsms}
Due to the low-star formation rate in UFDs, it is reasonable to ask whether any NS-NS or BH-NS mergers are expected in the population of observed UFDs. Observations indicate that the total star formation history of all 10 UFDs studied by \citet{2015arXiv151201558J} amounts to only 1.0$\times$10$^5$~M$_\odot$~\citep{Bechtol:2015cbp, McConnachie:2012vd}. While mass loss and tidal disruption may distort these measurements, we note that these observations lie on the mass-metallicity relationship, indicating that the mass lost from these systems has not drastically affected their stellar populations since the onset of their first supernovae~\citep{Kirby:2008ab}. In order to calculate the total number of NS-NS and NS-BH binary progenitors in UFDs, we employ a Kroupa ~\citep{Kroupa:2003jm} initial mass function with a minimum stellar mass of 0.08~M$_\odot$ and a high-mass index of $\alpha$~=~2.7, and conservatively assume a binary fraction of unity, with a flat secondary mass distribution~\citep{2006astro.ph..5069K}. Using these values we find that the UFD population could form $\sim$800 initial binary systems with an initial primary mass exceeding 5~M$_\odot$.

\begin{table}
\begin{tabular}{ c | c c c c c c c}
\hline \hline
- & 10 Myr & 50 Myr & 100 Myr & 500 Myr & 1 Gyr & 10 Gyr\\
\hline
10 km/s  & $<$0.0001 & $<$0.0001 & $<$0.0001 & 0.0011 & 0.0016 & 0.0023\\
20 km/s  & $<$0.0001 & 0.0004 & 0.0008 & 0.0085 & 0.0125 & 0.0183\\
50 km/s  & $<$0.0001 & 0.0064 & 0.0136 & 0.0569 & 0.0801 & 0.1345\\
100 km/s & 0.0002 & 0.0151 & 0.0378 & 0.1519 & 0.2202 & 0.4497\\
\hline
\end{tabular}
\caption{\label{tab:dnsrate} 
\begin{justify}
The Number of Expected NS-NS and NS-BH mergers within the population of UFDs considered in \citet{2015arXiv151201558J}, for various values of the UFD escape velocity, and the maximum age of the NS-NS or NS-BH merger necessary to produce r-process enrichment in the Reticulum II dwarf. We note that the constraints in Section~\ref{sec:nsms} argue strongly for a maximum kick velocity $\sim$10~km~s$^{-1}$ and an age $\sim$100~Myr, which results in the production of virtually no NS-NS or NS-BH mergers. The remainder of the columns shown in this table do not illustrate reasonable parameter space choices, but merely illustrate the robustness of our result to corybantic variations.
\end{justify}
}
\end{table}

To calculate the number of NS-NS mergers produced by this ensemble of systems, we utilize the results of~\citet{Dominik:2012kk} , noting that the modeled results are produced at a higher metallicity of [Fe/H]~=~-1.0, and that this choice is conservative given that NS-NS mergers are similarly efficient at both metallicities~\citep{Dominik:2013tma}. We have confirmed this result using a simulation of 2.0$\times$10$^{6}$ binaries at a metallicity [Fe/H]~=~-2.30 finding the results to be identical to within the level of Poisson noise\footnote{We acknowledge the Synthetic Universe Project (www.syntheticuniverse.org) for making the intermediate data products of \citet{Dominik:2012kk} publicly available for usage in these calculations, as well as Michal Dominik and Chris Belczynski for providing the low-metallicity dataset from \citet{Dominik:2013tma}.}. In what follows we quote the statistical results for simulations at higher metallicity, as the larger number of test systems decreased the Poisson noise for these rare events. We remain agnostic as to the best models of binary stellar evolution and marginalize our results over all 16 models presented in ~\citet{Dominik:2012kk}. Conservatively (realistically) assuming that the NS-NS or NS-BH merger must occur within 1~Gyr (100 Myr) from system formation to produce r-process materials before the formation of the remaining UFD stellar population, we find that 2.1 (1.5)  NS-NS and NS-BH mergers would be expected among the UFD population, in line with expectations.

However, the progenitor compact objects in a NS-NS and NS-BH merger must undergo significant natal kicks to move the system into a tightly bound, eccentric orbit~\citep{Willems:2003vb}. At the time of reionization, the progenitors of observed UFDs are expected to have masses $\sim$10$^7$~M$_\odot$~\citep{Salvadori:2008ju, 2015ApJ...799L..21W, Bland-Hawthorn:2015fua}, and escape velocities given by~\citep{Bovill:2010by}:

\begin{equation}
v_{esc}~=~10.9 \left(\frac{M}{10^7 M_\odot}\right)^{1/3} \left[\frac{1+z}{9.5}\right]^{1/2} \frac{\textrm{km}}{\textrm{s}}
\end{equation}

Examining the models of \citet{Dominik:2012kk} and removing binaries that received a center of mass kick (during either compact object formation event) larger than 10 km/s (20 km/s), we find that the total UFD population would be expected to produce $<$0.0001 (0.0008) mergers within 100~Myr after binary formation, and 0.0016 (0.0125) mergers within 1~Gyr after binary formation. In Table~\ref{tab:dnsrate} we summarize these results, including several additional models for both the maximum merger age and escape velocity for the UFD population. We note that Reticulum II may have a unique star formation history, DM halo formation history, or escape velocity, compared to other UFDs. In case Reticulum II has special properties compared to other UFDs, we recalculated the number of expected NS mergers, based on the 2.6$\times$10$^3$~M$_\odot$ of star formation observed in Reticulum II. With this smaller stellar population, we find that r-process events in Reticulum II from a NS merger becomes less likely. 

\section{Alternative Astrophysical Models}
\label{sec:alts}

In the prior section we have considered the merger of binaries, occurring through the joint stellar evolution of two, initially bound stars. However, alternative scenarios are possible. For example, in globular clusters, the majority of stellar encounters are believed to occur through n-body dynamics, due to the high stellar density in globular cluster centers. However, UFDs, are expected to be underdense (compared to globular clusters) by nearly six orders of magnitude, making stellar encounters negligible. A more important consideration, is whether NS-NS and NS-BH mergers in UFDs could be enhanced due to Kozai oscillations in triple systems~\citep{Thompson:2010dp, Sharpee:2012xj}. Notably, the eccentricities induced in the binary orbit through the Kozai mechanism can vastly decrease the merger time between widely separated binary companions. If we remove our cuts on the binary merger time in the previous section (assuming that the Kozai Mechanism produces a binary merger within the required 100~Myr timeframe regardless of the initial binary configuration) then we would produce 1.7 (2.0) NS-NS and NS-BH binaries with maximum kick velocities of 10 km/s (20 km/s). However, the fraction of triples in UFDs is unlikely to be unity, and it may also be difficult to keep these widely separated, low-mass, triple systems bound due to the mass loss and natal kicks of each supernova. A full investigation of this effect is warranted, but lies beyond the scope of this paper.

A second possibility includes r-process enrichment from only a small subset of supernova events, such as those that produce a rapidly spinning, high-magnetic field, magnetar~\citep{2008ApJ...676.1130M,Nishimura:2015nca,2015ApJ...811L..10T}. While none have been observed, they are expected to be short lived, and so may have escaped detection thus far. Since the rate of rapidly rotating magnetars can be adjusted freely, it is difficult to utilize stochasticity arguments to determine whether this mechanism could produce a single r-process event throughout the population of UFDs.

Some authors have argued for r-process enrichment connected to the accretion induced collapse of a white dwarf into a NS~\citep{1992ApJ...391..228W}. While this method is connected to binary dynamics, its rate, dynamical equation of state, and r-process yield are unknown~\citep{Fryer:1998jb,Thompson:2001ys,Piro:2014kfa}. However, similar to core collapse supernovae, individual accretion induced collapse events imply r-process abundance scales linearly with star formation rate, and thus metallicity, while Reticulum II and other observed UFDs are metal poor.

It should also be considered whether r-process enrichment of Reticulum II might result from direct, accretion-induced collapse of a NS, an event that should be significantly rarer than the AIC of white dwarfs. AIC of a neutron star would result after either a high-mass or low-mass X-Ray binary phase with mass transfer rates near the Eddington limit, which is $\sim$10$^{-8}$~M$_\odot$~yr$^{-1}$ for a NS. Assuming a maximum NS mass of 2~M$_\odot$ and an initial NS mass of 1.4~M$_\odot$, this corresponds to persistent Eddington limited accretion over the course of 60~Myr. While some NS systems (e.g. the recently discovered M82 X-2~\citep{2014Natur.514..202B}), may produce X-Ray luminosities that significantly exceed the Eddington limit, these systems are thought to be transient, with the majority of observed NS X-Ray binaries having maximum luminosities no more than a factor of two above the Eddington Limit~\citep{2003MNRAS.339..793G}. In the case of high-mass X-Ray binaries, model runs performed at low metallicity (Z~=~0.02Z$_\odot$) predict only 0.03 NS X-Ray binaries which produce consistent emission approaching the Eddington Luminosity for a starburst of the size expected in all observed UFDs -- 10$^{5}$~M$_\odot$~\citep{Linden:2010jx}. Systems with significantly brighter emission, such as M82-X2 have stable mass transfer periods limited to $\sim$10$^{5}$~yr~\citep{Fragos:2015vta}. In the case of low-mass X-Ray binaries, Roche Lobe overflow systems may persist on much longer timescales. Constraining ourselves to systems which transfer 0.6~M$_\odot$ within 500~Myr, we consider any persistent system with an X-Ray luminosity above 0.1L$_{edd}$, or approximately 1$\times$10$^{37}$~erg~s$^{-1}$. Studies by \citep{2008ApJ...683..346F} indicate that approximately 1.7$\times$10$^{-4}$ persistent LMXBs, with luminosities exceeding 10$^{37}$~erg~s$^{-1}$, would be produced in a 10$^{5}$~M$_\odot$ system. While this model assumes higher metallicity and continuous star formation, this is conservative for our calculation, as low-mass X-Ray binaries are extremely long-lived, and many form after the 500~Myr simulation time necessary to produce r-process enrichment in our model. Additionally, observations of globular clusters indicate that low-mass X-Ray binary formation is stronger in higher metallicity systems than at low-metallicities~\citep{2013ApJ...764...98K}. Altogether, it seems r-process enrichment of Reticulum II via direct collapse of a NS is disfavored.

A final possibility is that a NS merger occurred in Reticulum II, but that the NS binary was sourced by another star forming region, such as the ancient stars of the MW bulge. Assuming a total star formation of 10$^{10}$~M$_\odot$ within the first Gyr after bulge formation~\citep{2009IAUS..258...11W}, we calculate an expected population of $\sim$4$\times$10$^5$ double NS mergers, the majority of which happen far from the MW bulge due to the natal kicks given to the NS population. Using the kick velocities, and binary merger times calculated in \citet{Dominik:2012kk}, we find only a probability of only 0.008\% that a binary merger could occur within 100~pc of Reticulum II, assuming an average separation between the MW center and Reticulum II of 20~kpc during that epoch. However, this number may dramatically increase if the Reticulum II dwarf resided close to the MW bulge during early periods of intense star formation.  

\section{The R-Process From Dark Matter}
\label{sec:darkrprocess}
The r-process abundance in Reticulum II and the MW, could result from dark matter instigating lone NS implosions that eject neutron-rich fluid. Extensive radio searches by the Green Banks Telescope and Arecibo array of the inner parsecs of the Milky Way galaxy, have not revealed the expected population of galactic center (GC) pulsars \citep{Macquart:2010vf,Wharton:2011dv,Dexter:2013xga,Chennamangalam:2013zja}. In \citet{Bramante:2014zca}, we found that DM could implode NSs in the GC and thus account for the missing pulsars. Dark matter, more dense in the central parsec of the Milky Way, can accumulate at the center of NSs and form a star-consuming BH within $t_{\rm c}\sim 10^5 - 10^8$ years, depending on the mass, local DM density ($\rho_{\rm DM}$), and nucleon scattering cross-section ($\sigma_{\rm nX}$) of the DM particle. 

Outside the center of the Milky Way, measurements of the characteristic age of pulsars, corroborated by the ages of binary partner white dwarfs, indicate that disk pulsars reach ages of at least $\sim {\rm Gyr}$ \citep{Bramante:2015dfa}. This puts an upper bound on $\sigma_{\rm nX}$ for a number of DM models \citep{Goldman:1989nd,Starkman:1990nj,Bertone:2007ae,Kouvaris:2007ay,deLavallaz:2010wp,Kouvaris:2010jy,McDermott:2011jp,Kouvaris:2011fi,Kouvaris:2011gb,Bramante:2013hn,Bell:2013xk,Bertoni:2013bsa,Bramante:2013nma,Guver:2012ba,Zheng:2014fya,Perez-Garcia:2014dra,Brito:2015yga,Kurita:2015vga}. 

Although NSs in the Milky Way disk may be as old as $10^{10}~{\rm yrs}$, NSs can still quickly implode in the Galactic center. The DM capture rate in NSs scales linearly with $\rho_{\rm DM}$ and $\sigma_{\rm nX}$. Therefore, because DM density at the center of the MW is up to $10^4$ times denser than in the disk, pulsars in the central parsecs can collapse after $\lesssim 10^6$ yrs, and are potentially sensitive to $10^4$ times larger $\sigma_{\rm nX}$ \citep{deLavallaz:2010wp,Bramante:2014zca}. During NS collapse, the reconfiguration of the NS's magnetosphere radiates $\sim 10^{42} ~{\rm erg}$ in a millisecond; hence DM-induced NS implosions could also be the source of fast radio bursts (FRB) \citep{Fuller:2014rza}. 

For concreteness, in the following calculation of the r-process yield from a DM induced NS implosion, we assume a non-annihilating, negligibly self-interacting DM candidate with a mass $m_{\rm X} = 10 ~{\rm PeV}$, that can either be bosonic or fermionic, and with a DM-nucleon scattering cross-section $\sigma_{\rm nX} \gtrsim 10^{-45}~{\rm cm^2}$. There are additional DM models, e.g. shift-symmetric bosons \citep{Bramante:2014zca} and Higgs portal fermions \citep{Bramante:2015dfa}, which would induce GC pulsar collapse for DM masses $m_{\rm X}={\rm keV-PeV}$. We consider PeV mass DM because its predicted cross-section complements imminent direct detection searches and it may be responsible for type Ia supernovae ignition \citep{Bramante:2015cua,Graham:2015apa}. We assume NSs with radius $R_{\rm NS} \sim 10~{\rm km} $, mass $M_{\rm NS} \sim 1.5~{\rm M_{\odot}}$, temperature $T_{\rm NS} \sim 10^4 ~{\rm K}$, and central density $\rho_{\rm NS} \sim 10^{15} {\rm ~g ~cm^{-3}}$. With these parameters, any $\sigma_{\rm nX}$ in excess of $10^{-45}~{\rm cm^2}$ saturates the geometric cross-section for DM capture in a NS. Therefore, $\sim$PeV mass DM-induced NS collapse models are robust against variations in $\sigma_{\rm nX}$. For details, see \citep{Bramante:2015cua}.

Hereafter we demonstrate that during the growth of a small BH inside a NS, $\sim 10^{-5}-10^{-1}~M_{\odot}$ neutron-rich fluid could be ejected as a result of tidal squeezing. To precisely determine the mass ejected, it would be necessary to perform a detailed hydrodynamic simulation of tidal forces during the (likely turbulent) process of a NS's rapid inward accumulation onto a growing black hole. We leave hydrodynamic simulations to future work. In the remainder of this section, we (1) find the maximum mass of r-process elements ejected, (2) show that the physical conditions prompting neutron fluid ejection in a NS-BH merger are also present when a NS implodes into a black hole, and (3) find that substantial mass ejection by neutrino emission must occur outside a steady-state regime.

To set an upper limit on the mass that escapes from an imploding NS, we specify the binding energy of the implosion, $E_{\rm i} \approx 3GM_{\rm NS}^2 (R_{Sch. \rm}^{-1} - R_{NS \rm}^{-1})/5  = 3 \times 10^{57} (M_{\rm NS} / 1.5 M_{\odot}) ~{\rm GeV} $, where $R_{Sch. \rm}$ is the star's Schwarzschild radius. We compare this binding energy to the energy required to accelerate a nucleon to escape velocity ($v_{\rm ej} \sim 0.7 ~{\rm c}$) at the surface of the NS, $E_{\rm a} = \gamma(v_{\rm ej}) m_{\rm n}  $, where $m_{\rm n}$ is the nucleon mass. The maximum mass of ejected material is
\begin{align}
M_{ej \rm} \leq m_{\rm n} \frac{E_{\rm i}}{E_{\rm a}} \lesssim 0.2 ~ \left( \frac{M_{\rm NS}}{1.5 M_{\odot}} \right) \left( \frac{1.4}{\gamma(v_{\rm ej})} \right)~M_{\odot}.
\label{eq:mejupperbound}
\end{align}

R-process production accompanies a NS-BH (or NS-NS) merger when ejected neutron-rich fluids decompress, providing enough free neutrons to synthesize heavy elements \citep{1977ApJ...213..225L,Tanaka:2013ixa,Bauswein:2014vfa}. The mechanism for neutron fluid expulsion from a NS as it approaches a black hole, first described in \citep{1971swng.conf..539W} as the "tube of toothpaste" effect, was developed in \citet{1973ApJ...185...43F,1975ApJ...197..705M,1976ApJ...210..549L}. As neutron fluid crosses the Roche limit of a black hole (or other compact body), the tidal squeeze from the black hole will eventually exceed the self-gravity of the neutron fluid. The resulting severe compression propels streams of neutron fluid away from the black hole. 

Here we verify that during dark-matter-induced NS implosions, the entire NS crosses the Roche limit as it flows into the black hole formed at its center. Initially, the Roche limit for the neutron fluid surrounding a newly formed BH is
\begin{align}
R_{\rm Roche}  
\simeq 20 \left(\frac{M_{\rm BH}}{10^{-10}~ {\rm M_{\odot}}} \right)^{1/3} \left(\frac{10^{14} ~{\rm g~cm^{-3}}}{\rho_{\rm NS}} \right)^{1/3}~{\rm m},
\end{align}
where $\sim 10^{-10} ~{\rm M_{\odot}}$ is the maximum DM mass a NS collects in 10 Gyr for $\rho_{\rm DM} \sim 10^4~{\rm GeV~cm^{-3}}$, see \citep{Bramante:2013hn}. As the BH grows, the Roche limit expands, passing through the entire NS by the time $M_{\rm BH} \sim 0.02~{\rm M_{\odot}}$.

To approximate the time for a natal black hole to grow to $M_{\rm BH} \sim 0.02~{\rm M_{\odot}}$, at which point it squeezes and consumes the bulk of the NS, we employ the Bondi accretion rate as in \citep{Markovic:1994bu,Kouvaris:2013kra}. The Bondi accretion rate of neutron fluid onto the black hole is
$
{\rm d} M_{\rm BH}/{\rm d} t = 4 \pi \lambda_s G^2 M_{\rm BH}^2 \rho_{\rm n}/v_{\rm s}^3,
$
where, in what follows, we take typical values for the neutron fluid's sound speed $v_{\rm s} \sim 0.3 ~{\rm c}$, density $\rho_{\rm n} \sim 10^{15} {\rm g/cm^3}$, and accretion constant, $\lambda = 0.25$, \citep{Shapiro:1983du}. Because the BH growth rate accelerates with added mass, the time for the BH to consume the remainder of the NS is set by its initial mass. For $m_{\rm X} = 10~{\rm PeV}$ mass DM, a BH forms once $\sim 10^{-14} ~{\rm M_\odot}$ DM has collected at the center of the NS, \citep{Colpi:1986ye,Bramante:2015cua}. This BH grows until the Roche limit encompasses the NS after
\begin{align}
t_{\rm imp} \sim \frac{v_{\rm s}^3}{4 \pi \lambda_s G^2 M_{\rm BH} \rho_{\rm n}} \simeq 50 \left( \frac{10^{-14} ~{\rm M_\odot}}{M_{\rm BH}} \right) ~{\rm yrs}.
\end{align}
This also implies that in a span of $\sim 1~ {\rm ms}$, the remaining $1.4~{\rm M_\odot}$ of neutron fluid crosses the Roche limit.

Finally, we note that because the bulk of the NS is accumulated within $\sim 1 ~{\rm ms}$, this disfavors the canonical r-process production mechanism proposed for core collapse supernovae \citep{Woosley:1994ux}, wherein an outpouring of neutrinos heat nuclear material expelled from the surface of a proto-neutron star \citep{1986ApJ...309..141D}. In the standard scenario, which assumes steady-state neutrino-driven mass ejection from a collapse-heated NS, the neutrinos and neutron star crust must be in thermal equilibrium, which is established after $\sim 1 ~{\rm s}$, too long compared to a $\sim 1 {\rm~ ms}$ timescale for the DM-induced NS implosion. 

\section{Dark R-Process Production in UFDs and the MW}
\label{sec:darkrates}

This section examines the  DM r-process enrichment of the Milky Way and UFDs. Results in the previous section indicate that each DM-induced NS implosion results in $1.5 ~{\rm M_{\odot}}$ of NS material crossing the Roche limit of a black hole. Recent simulations of NS-BH mergers find that $\sim10^{-4} - 10^{-1} ~{\rm M_{\odot}}$ of ejecta is expelled at $\sim 0.1-0.3 ~{\rm c}$ as NSs cross the Roche limit of $3-10~{\rm M_{\odot}}$ mass black holes, with lower mass and higher spin BHs tending to yield higher mass ejecta, \citep[see e.g.][]{Shibata:2011jka,Foucart:2012vn,Deaton:2013sla,Bauswein:2014vfa,Kyutoku:2015gda}. Hereafter, we show that MW and UFD r-process enrichment from DM-induced implosions, favors an ejecta mass $M_{\rm ej} \sim 10^{-5} -10^{-3}  ~{\rm M_{\odot}}$.

\begin{figure}
\includegraphics[width=0.47\textwidth]{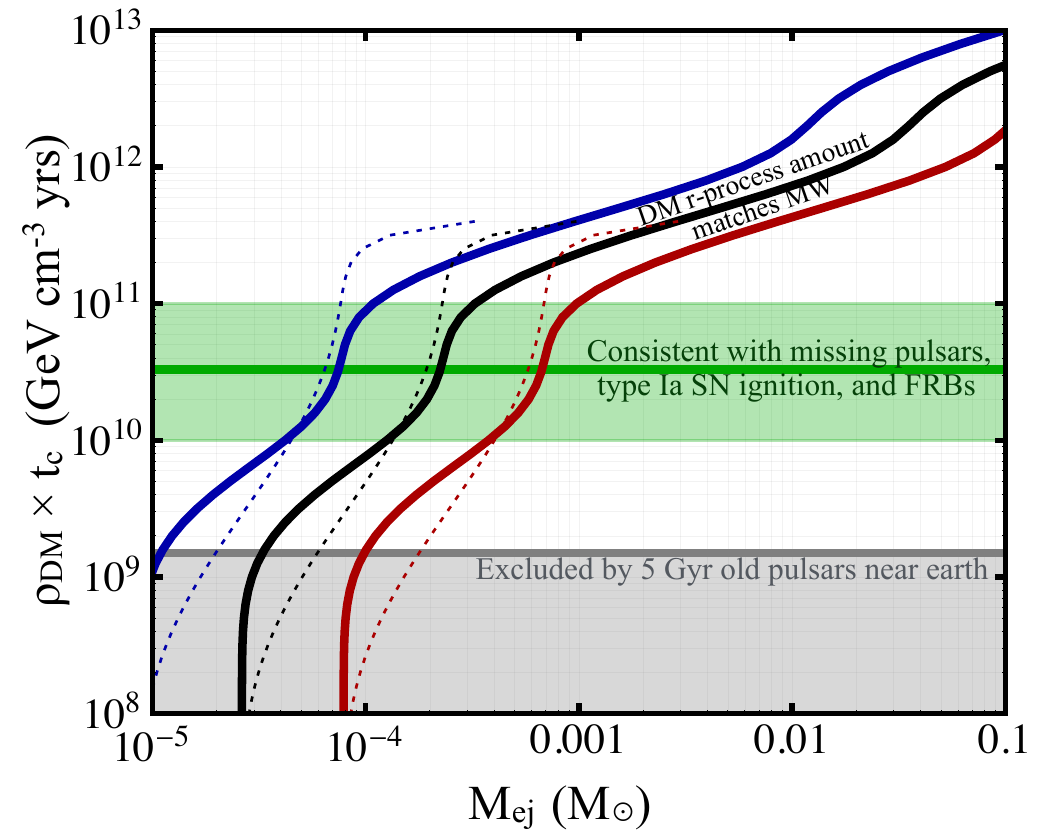}
\caption{The y-axis is the integrated time and local DM density required to implode a NS, i.e., a NS surrounded by DM density $\rho_{\rm DM} $ implodes after time $t_{\rm c}$ . The x-axis indicates the average r-process mass ejected in a NS implosion. The green band indicates parameter space favored for the missing pulsar problem, GC implosions as FRB candidates, and DM ignition of type Ia supernovae. The grey region is exluded by old pulsars found near earth. The thick black line shows parameter space where NS implosions in the MW GC provide $10^4~M_{\odot}$ of r-process elements, consistent with observations, for the NFW DM profile detailed in the text. The thick red and blue lines show a factor of three variation in total Milky Way r-process mass production for an NFW DM profile. The dotted lines show the same parameter space, where $10^4~M_{\odot}$ of r-process elements are produced, but assume a Burkert dark matter density profile, with a constant dark matter density, $\rho_{DM}=50~{\rm GeV/cm^3}$, inside the central kiloparsec of the Milky Way. These lines truncate at $\rho_{\rm DM} t_c \sim 5 \times 10^{11} ~{\rm GeV~yrs}$, above which neutron stars in the MW will not implode for a maximum MW DM density of $\rho_{DM}=50~{\rm GeV/cm^3}$.}\label{fig:rhomej}
\end{figure}

Studies of the production and hydrodynamic redistribution of r-process elements in the Milky Way \citep{Argast:2003he}, have recently been significantly improved to account for the time-evolved migration of r-process elemental abundances, including the affects of galactic subsystem mergers, mixing in the interstellar medium, and (outflows from) star formation during the dynamical evolution of the Milky Way \citep{2015MNRAS.447..140V,2015ApJ...807..115S,Wehmeyer:2015sra}. Particularly, it has been shown that, as long as r-process production events occur at sites where neutron stars have formed within the inner kpc of the Milky Way, models will match the observed r-process abundance and scatter in [Fe/H] versus [Eu/Fe] to presently available precision. The work of \citep{2015MNRAS.447..140V} demonstrated that r-process chemical evolution is consistent with observations so long as two criteria were met by an underlying neutron star-sourced r-process production model: (1) the time for r-process production to occur after neutron star formation should fall within $t_{min} \sim 3\times 10^6-3\times 10^8 ~{\rm yrs}$, (2) the amount of r-process elements produced should be $\sim 10^4$ solar masses, implying a rate between $10^{-5}$ and $10^{-6}$ r-process events per solar mass of stars in the Milky Way. Building on these results, we consider the scenario of neutron star implosion-induced r-process production in the MW to be viable, so long as it matches the per stellar mass rate ($10^{-5}-10^{-6}~{M_\odot}^{-1}$), total r-process abundance ($\sim 10^{4} M_\odot$ total r-process materials in the MW), and minimum time for r-process enrichment to begin ($t_{min} < 10^{8}~{\rm yrs}$) found in \citep{2015MNRAS.447..140V}. 

In order to obtain the number of neutron stars which formed close enough to the center of the galaxy that they could have already imploded (compared to longer-lived NSs outside the galactic center), we model the stellar density of the inner $\sim 30$ kpc of the MW using velocity-curve-fitted density parameters provided in \citep{Sofue:2013kja}, namely $\rho_{\rm *}(r) = \sum_{i=1}^3 \rho_i e^{-r/a_i}$, where ($\rho_1=4\times 10^4,~\rho_2=2\times 10^2,~\rho_3=0.1{\rm ~~M_\odot~pc^{-3}}$) and ($a_1=0.0038,~a_2=0.12,~a_3=3{~~ \rm kpc}$). One small difference between the neutron star implosion scenario and the neutron star merger scenario, is that, for parameters that match the total observed r-process abundance, NS-implosions will occur \emph{mostly} within the central kpc, where the DM density is $\gtrsim 100$ greater and neutron star implosions can occur. However, neutron star mergers are also expected to occur \emph{mostly} within the central few kpc, so we do not expect the subsequent chemical and hydrodynamical evolution of r-process enrichment in the NS implosion scenario to differ substantially from the neutron star merger scenario. This conclusion is supported by simulations, showing that after r-process elements are produced, they are mixed through the outer $\sim$30 kpc of the MW on $\lesssim$ Gyr timescales \citep{2015MNRAS.447..140V,2015ApJ...807..115S}.

As previously explained, the observed abundance and scatter of r-process elements observed in the MW require $\sim 10^4  ~{\rm M_{\odot}}$ of r-process production over $\sim 10^{10}$ yrs, with a substantial fraction produced within $\sim 10^7$~yrs of the initial star formation epoch. The second requirement is fulfilled naturally by GC NS implosions, which often occur within $\sim 10^7$ yrs in the inner kiloparsec, for DM parameters that solve the missing pulsar problem (the parameter space shown within the green band in Figure \ref{fig:rhomej}). In Figure \ref{fig:rhomej} we show DM-NS-implosion parameters for which a per-NS-implosion ejecta mass $M_{\rm ej}$ would provide $10^{4} ~M_{\odot}$ of r-process elements over the lifetime of the MW, also fulfilling the requirement that the first implosions occur within $10^7$ yrs. The thick lines in Figure \ref{fig:rhomej} show parameter space, where the fraction of NSs formed in the inner kpc, that collapse within $10^{10}$ yrs provide for the MW's total r-process abundance. These thick lines have been made by using the aforementioned Sofue stellar distribution profile, along with an NFW dark matter halo profile, $\rho_{\rm DM}^{\rm NFW} (r) = \rho_0/((r/r_{\rm s})(1+r/r_{\rm s})^2)$, where we take $\rho_0 = 0.24 ~{\rm GeV~cm^{-3}}$ and $r_{\rm s}=20~{\rm kpc}$. To find the number of neutron stars contained within radius $r$, we use the ~~\citet{Kroupa:2003jm} initial mass function as in Section~\ref{sec:nsms} and a minimum zero-age main sequence mass of 8~M$_\odot$, normalizing with the assumption that $\sim$0.5\% of $\gtrsim8~M_\odot$ stars formed in the central 10~kpc will evolve to form NSs. Assuming a constant $200~{\rm km/s}$ MW DM velocity dispersion, we calculate the number of neutron stars formed close enough to the galactic center, that they will implode within the lifetime of the Milky Way.

To demonstrate that these results are robust against variations in the assumed DM halo density profile, we also employ a Burkert dark matter halo profile \citep{Burkert:1995yz}, $\rho_{\rm DM}^{\rm Burk} (r) = \rho_0^B/((1+r/r_{\rm s,B})(1+r^2/r_{\rm s,B}^2))$, where we take $\rho_0^B = 50~{\rm GeV/cm^3}$ and $r_{\rm s,B}=1~{\rm kpc}$. The Burkert profile is flat or ``cored" at its center, and with the aforesaid parameters, DM will have a constant $50~{\rm GeV/cm^3}$ density inside the central kpc of the Milky Way. Figure \ref{fig:rhomej} shows that in preferred NS-implosion parameter space, this does not substantially change the result.

\begin{figure}
\includegraphics[width=0.47\textwidth]{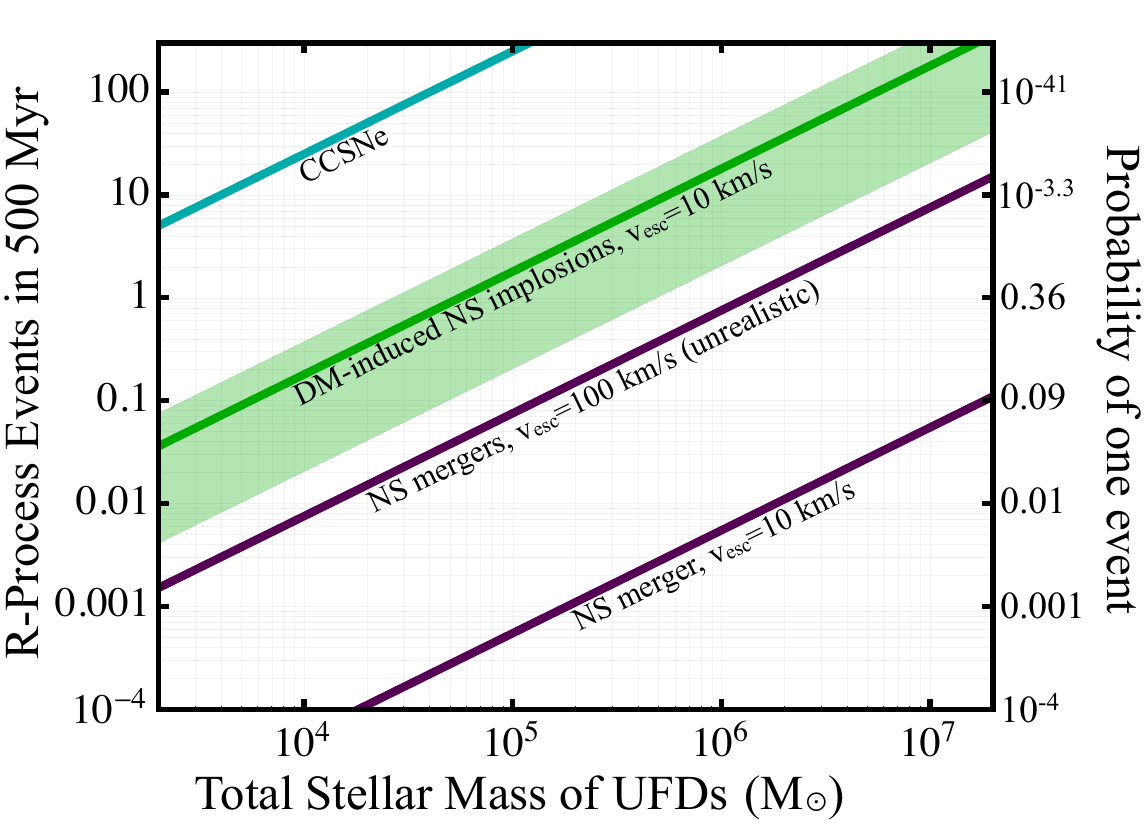}
\caption{The number of r-process events expected after 500 Myr in the ensemble of UFDs analyzed in \cite{2015arXiv151201558J}, assuming half of the total stars form in a burst at the birth of the UFDs. The green line and band match the DM-NS-implosion line and band shown in Figure \ref{fig:rhomej}. The extremely high density of r-process elements in Reticulum II compared to other UFDs favors $\mathcal{O}(1)$ r-process events integrated over this population.\label{fig:RUFDrates}}
\end{figure}

In Figure \ref{fig:RUFDrates} we display the expected rate of NS implosions in UFDs analyzed in \citep{2015arXiv151201558J}. The star formation histories of UFDs are an area of active research; in accord with the results of \citep{2014ApJ...796...91B,Bland-Hawthorn:2015fua}, we show the number of r-process events after 500 Myr, assuming half the UFD stars formed in a burst at $z\sim 5$. We take a Plummer stellar density in UFDs, $\rho_{\rm p}(r) = (3 M_{\rm UFD}/4\pi b^3)(1+r^2/b^2)^{-5/2}$, where $b\sim 42~{\rm pc}$. We assume the same 0.5\% NS formation fraction as in the MW, and using the double exponential NS kick model in \citep{FaucherGiguere:2005ny}, we find that 2\% of NSs formed inside a UFD will experience a natal kick $<$5~km/s, and so remain bound inside the central parsecs. Note that a single event which produces $\sim 10^{-3}-10^{-4}~{\rm M_{\odot}}$ r-process materials within $500$ Myr in the ten surveyed ultra-faint dwarf spheroidals (along with Reticulum II) is consistent with the observed r-process abundance in these systems \citep{2016Natur.531..610J,2016AJ....151...82R}.

We then find the radius within which NSs implode within 500 Myr in the UFD. The UFD DM density is assumed to follow an NFW profile as defined above, with $\rho_0=10^{-2}~{\rm GeV/cm^3}$ and scale factor $r_{\rm s}=1.5~{\rm kpc}$, in accord with a $10^7~{\rm M_{\odot}}$ DM halo at redshift $z\sim 5$ \citep{Klypin:2010qw}. For fixed DM density, DM capture by NSs in the central 50 parsecs of UFDs will be $\sim 200$ times greater than in the MW, because DM capture in NSs scales inversely with halo velocity dispersion, which is about $1~{\rm km/s}$ in the central 50 parsecs of the UFD at $z\sim 5$. Figure \ref{fig:RUFDrates} shows that for UFDs with the aforestated stellar and DM density profile, the DM-induced NS implosion model predicts $\mathcal{O}(1)$ event within 500 Myr, in line with observation of high r-process abundance in Reticulum II, given the stochasticity of r-process enrichment in the UFD population. We also calculated the expected rate for a Burkert DM profile, with $\rho_{0,B}=10^{-2}~{\rm GeV/cm^3}$ and scale factor $r_{\rm s,B}=50~{\rm pc}$. As with the r-process enrichment of the Milky Way, this changed the result by less than a factor of two.

\section{r-process enrichment of UFDs vs. Globular Clusters}
\label{sec:rglob}

One potential diagnostic that can differentiate baryonic and DM mechanisms of r-process production is r-process enrichment in globular cluster populations compared to that in UFDs. Both environments exhibit simple star formation histories, dominated by a single massive star formation event, and contain stars with extremely low and uniform metallicities. The systems differ in two important aspects: (1) the DM density in the center of UFDs is extremely high, (2) the stellar density of globular clusters is orders of magnitude higher than UFDs, significantly enhancing the binary (and ternary) stellar processes which lie behind the majority of proposed baryonic r-process mechanisms. 

Intriguingly, \citet{Roederer:2011id} studied 11 globular clusters and found no stars with an r-process enrichment [Eu/Fe]~$>$~1.2, a result that agrees with earlier studies of NGC 6397, NGC 6752 and 47 Tuc, which found average [Eu/Fe] ratios below 0.47~\citep{2004A&A...427..825J}.  On the other hand, six of the nine stars observed in Reticulum II contain [Eu/Fe] ratios exceeding 1.68~\citep{2015arXiv151201558J, 2016arXiv160104070R}. The preference for significant r-process enrichment in Reticulum II (compared to any globular cluster) is difficult to explain in terms of any binary (or ternary) stellar mechanism, since the rate of such encounters is expected to be orders of magnitude higher in the globular cluster population. If upcoming observations discover more UFDs with r-process enrichment, and more extensive surveys of globular clusters find no such enrichment, this would further support a DM origin for the r-process in UFDs. 

\section{Conclusion}
\label{sec:conc}
We have shown that NS mergers are unlikely to produce the r-process overabundance observed in the Reticulum II dwarf, since the total production rate of NS mergers is low, and supernova natal kicks efficiently remove binary stellar systems from the shallow gravitational well of UFDs. Additionally, we have examined several alternative explanations, finding that Kozai oscillations in a ternary system or r-process production in quickly rotating magnetars could potentially explain the observed signal, while the remaining models for r-process production appear in tension with the data. 

We have found that DM-induced NS implosions could be the source of r-process enrichment in UFDs and the MW. This establishes a connection between the high DM density in UFDs and the r-process enrichment observed in Reticulum II. 
In addition, the DM-induced r-process scenario predicts that ejecta from DM-induced NS collapse will power an FRB kilonova afterglow, akin to the signal expected from NS-NS merger kilonovae \citep{Li:1998bw,Metzger:2010sy,Goriely:2011vg,Kasen:2013xka}. Our findings also imply an r-process signature for primordial black hole capture on NSs \citep{Abramowicz:2008df,Capela:2013yf,Pani:2014rca,Defillon:2014wla,Graham:2015apa}, and in turn, small scale primordial perturbations \citep{Carr:1994ar}. Of course, as with other proposed r-process sites (i.e. core collapse supernovae, swiftly rotating magnetars, and neutron star mergers), DM-induced NS implosions may also provide a fraction of the total r-process elements in the Milky Way.

It is worth noting that Reticulum II is now unique among UFDs in two independent ways: (1) it is the only UFD providing a possible $\gamma$-ray signal, and (2) it is the only UFD showing signs of r-process enrichment. In the case that r-process production is caused through the DM induced collapse of single NSs, these two observations are both consistent with enhanced DM density in the central region of the Reticulum II dwarf. Further observations will be necessary to rule out, or confirm, the contribution of DM to both the $\gamma$-ray signal and the r-process enrichment of Reticulum II. The results of these studies have the potential to provide new insights into the nature of DM.

\vspace{3mm}
\noindent{\bf Acknowledgements: } 

\noindent We thank John Beacom, Keith Bechtol, Chris Belczynski, Michal Dominik, Alex Drlica-Wagner, Jennifer Johnson, Vicky Kalogera, Paul Lasky, Adam Martin, Grant Mathews, Annika Peter, Kris Sigurdson, Rebecca Surman, Todd Thompson, and MacKenzie Warren for helpful discussions, and the ABHM Workshop for their hospitality while this work was completed. JB thanks the Aspen Center for Physics, which is supported by National Science Foundation grant PHY-1066293. TL is supported by the National Aeronautics and Space Administration through Einstein Postdoctoral Fellowship Award No. PF3-140110. 

\bibliographystyle{apj}

\bibliography{dm_rprocess}

\end{document}